# Shifting Research Funding Priorities under Geopolitical Pressure: Evidence from Estonia

Yunfeng Gao [1], Yang Ding [2,*]

[1] School of Economics and Management, East China Normal University, Shanghai 200062, China

[2] Business School, The University of Edinburgh, Edinburgh EH8 9JS, United Kingdom

* Corresponding Author: Yang Ding (yang.ding@ed.ac.uk)

**Abstract**

This study examines how the 2014 Donbas-war breakpoint was associated with changes in the semantic composition of research funding recorded in Estonia, a geopolitically exposed country outside the belligerent states. It combines 17,952 research-funding records from the Estonian Research Information System (ETIS) for 2000 to 2019 with a field-by-year matched OpenAlex reference corpus and geocoded Donbas conflict records. Contrastive text projections distinguish explicit-war language from dual-use technological orientation and eight crisis-relevant capability channels, allowing shifts to be detected beyond projects that directly mention war or security. Segmented annual models show positive post-2014 slope changes in dual-use and maximum-capability measures, whereas explicit-war language follows a different trajectory. Financing-weighted estimates identify positive post-2014 level differences among larger recorded projects. Capability-channel comparisons indicate the strongest positive period differences in computer science, energy, and engineering, together with a temporary narrowing of the capability profile. The results suggest that a neighboring research system can register geopolitical pressure through changes in the technical and preparedness-oriented language of funded projects. More broadly, the study demonstrates how project-level semantic evidence can reveal shifts in public science-funding priorities that remain difficult to observe through disciplinary classifications or explicit conflict terminology alone.

**Acknowledgements:** Yang Ding acknowledges a PhD fellowship (no. s2222886) from the University of Edinburgh Business School.

# 1 Introduction

The twenty-first century has been marked by persistent geopolitical instability and recurrent regional wars. Armed conflict damages universities, laboratories, and scientific careers within belligerent states, while sanctions, migration, supply-chain disruption, and altered security demand can transmit effects across national borders (Caldara & Iacoviello, 2022; European Commission, 2022; Ganguli, 2017; Kuzhabekova, 2024; Waldinger, 2016). Research systems in neighboring countries may consequently reassess which technological and societal capabilities merit support even when they are not parties to the conflict. This study therefore asks a broad science-policy question: how does a nearby regional war become visible in the semantic direction of research funding in a geopolitically exposed neighboring country?

Research funding is a principal channel through which governments connect public problems with scientific activity. Mission-oriented policy directs resources toward socially defined challenges and coordinates knowledge production across agencies, universities, and firms (Foray et al., 2012; Kattel & Mazzucato, 2018; Mazzucato, 2018; Schot & Steinmueller, 2018). Defense-related research and development (R&D) further shows how operational requirements can generate research agendas in communications, sensing, materials, energy, health, and logistics without requiring every project to use overt military terminology (Moretti et al., 2025; Mowery, 2012). Yet this literature does not provide direct evidence about neighboring countries exposed to the geopolitical consequences of war.

The effects of public R&D depend partly on programme design and the technological fields that receive support. US National Institutes of Health (NIH) funding rules influence downstream private-sector patenting, defense R&D generates private and international knowledge spillovers, and mission programmes can leave durable institutional capabilities (Azoulay et al., 2019; Gross & Sampat, 2023; Howell et al., 2025; Moretti et al., 2025). These studies establish that the direction of public research can shape later innovation, but they focus mainly on funded outputs and direct programme effects. The earlier stage at which geopolitical pressure enters the language and orientation of project portfolios remains less well understood.

Geopolitical pressure may influence project language through several pathways. Researchers can connect emerging threats to accumulated expertise, programme managers can reinterpret existing capabilities in relation to preparedness, and funders can emphasize applications that bridge civilian and security needs. Such adjustment is likely to appear first in technical combinations and capability language because researchers commonly move toward topics adjacent to their established knowledge base (Hill et al., 2025). A semantic approach is therefore well suited to detecting changes that conventional disciplinary labels or counts of explicit war terms may overlook.

The Donbas war offers a strategically important setting for examining regional scientific adaptation before the escalation of the Russia-Ukraine war in 2022. Beginning in 2014, the conflict altered the security environment of Northern and Eastern Europe through renewed territorial warfare, cyber concerns, energy vulnerability, and heightened attention to resilience (European Commission, 2022). Estonia is geographically separated from the Donbas theatre, yet it shares a border and a long security relationship with Russia, belongs to the North Atlantic Treaty Organization (NATO) and the European Union, and had developed substantial cyber-security institutions following the 2007 cyberattacks (Crandall, 2014; Crandall & Allan, 2015; Wrange & Bengtsson, 2019). These features make Estonia an analytically informative neighboring case in which indirect geopolitical exposure can be studied within a stable research-information system.

Estonia also provides unusually detailed project metadata across the full 2000 to 2019 observation window. ETIS records project identifiers, titles, abstracts, research fields, recipients, timing, and reported financing for projects involving Estonian organizations. The long pre-2014 record supports comparison of the post-2014 pattern with more than a decade of earlier semantic development. The availability of project text and financing information also permits the analysis to distinguish the typical project, the upper tail of the semantic distribution, and the orientation of larger recorded projects.

This paper asks whether the onset of the Donbas war was associated with changes in the semantic direction of research funding recorded in Estonia. It links ETIS project metadata to the Uppsala Conflict Data Program Georeferenced Event Dataset (UCDP GED) and Armed Conflict Location & Event Data (ACLED) records and to a matched OpenAlex corpus that separates war and crisis scholarship from comparable nonwar research. The design distinguishes explicit-war language, dual-use technology, and eight capability channels covering engineering, materials, computer science, energy, medicine, life science, earth and environment, and agriculture and food systems. It then examines whether post-2014 change occurred in annual means, medians, upper-tail shares, financing-weighted scores, and the balance of the capability profile.

The study makes three substantive contributions. First, it shifts attention from scientific disruption inside belligerent states to the semantic adaptation of research funding in a neighboring country exposed to geopolitical change. Second, it distinguishes overt war language from dual-use and crisis-capability language, thereby identifying technical orientations that would remain hidden in a narrow security-keyword count. Third, it connects project-level text, award-size weighting, and capability-channel balance within a single longitudinal framework, allowing the location and composition of semantic change to be examined together.

The empirical design combines project-level semantic projections with annual segmented models centered on 2014. Means, medians, upper-tail shares, and financing-weighted scores show whether change is broad, concentrated among unusually aligned projects, or especially visible among larger recorded projects. Channel-specific measures identify where capability-oriented language is most pronounced, while annual entropy summarizes whether attention is distributed evenly across the eight capability domains. Together, these measures trace how a regional-security breakpoint appears in the direction and internal structure of a national research portfolio.

# 2 Data and methods

## 2.1 Research design and funding data

The primary data source is the Estonian Research Information System (ETIS), owned by the Estonian Ministry of Education and Research and managed by the Estonian Research Council. The analytical file contains 17,952 research-funding project records with a project year from 2000 to 2019. It includes national, European, institutional, private, and other recorded funding sources, so the empirical object is the full ETIS-recorded research-funding portfolio. **Appendix A** documents field harmonization, duplicate control, date construction, text availability, and financing coverage.

The 2000 to 2013 baseline contains 11,555 records and the 2014 to 2019 period contains 6,397. Project counts and the seven descriptive ETIS fields are summarized in Figure 1, which establishes the temporal and disciplinary support of the analytical sample. The figure also shows why annual semantic estimates are adjusted for observable metadata composition.

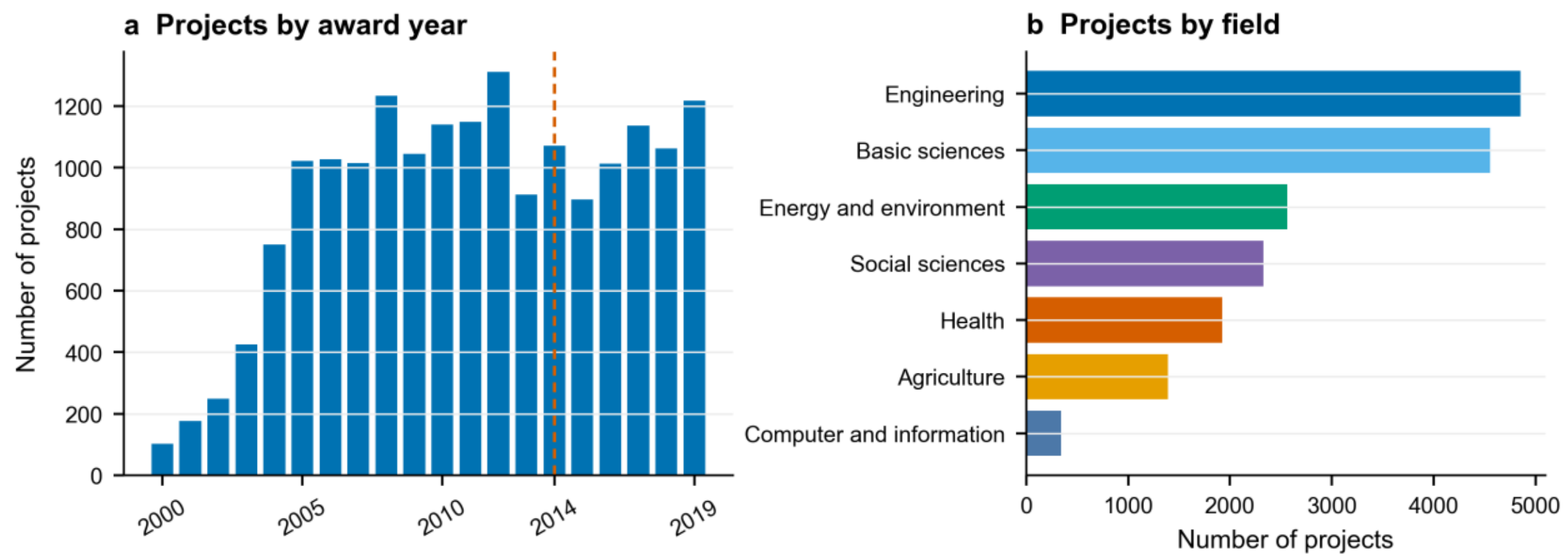

**Figure 1. Distributions of ETIS-recorded projects by year and descriptive field.** Both panels use the 17,952 ETIS research-funding records in the analytical file. Panel a reports records by project year and marks 2014. Panel b reports frequencies under the seven-category descriptive mapping documented in **Appendix A**.

Figure 1 shows a pronounced expansion in annual ETIS coverage during the early 2000s followed by a denser and more stable record stream after 2004. Engineering and basic sciences form the largest descriptive groups, with substantial representation from energy and environment, social sciences, health, agriculture, and computer and information research. This composition provides broad technical and societal coverage for the semantic analysis rather than a portfolio dominated by a single ETIS field. As shown in Table 1, ETIS supplies the target research-funding records, OpenAlex supplies the contrastive semantic reference, and the two conflict databases supply the regional-security context.

**Table 1. Data sources and analytical roles.**

| Data component | Coverage | Observations | Analytical role |
|---|---|---|---|
| ETIS-recorded projects | 2000–2019 | 17,952 records | Project and annual semantic outcomes |
| Pre-2014 records | 2000–2013 | 11,555 records | Baseline levels and trends |
| Post-2014 records | 2014–2019 | 6,397 records | Level and slope differences |
| War/crisis reference corpus | 2000–2024, field-year paired | 5,449 abstracts | Positive semantic poles |
| Nonwar comparison corpus | 2000–2024, field-year paired | 5,449 abstracts | Comparison semantic poles |
| Donbas conflict records | 2014–2019 | 30,104 records | Context and exploratory diagnostics |

Note: An ETIS record is the basic observation for semantic measurement. The project file includes multiple funding-source types and is therefore described as an ETIS-recorded portfolio. OpenAlex counts refer to the operative paired corpus used by the TF-IDF scoring script. Conflict records are used for context and exploratory alignment tests.

## 2.2 Conflict exposure and external semantic corpus

Conflict context combines the Uppsala Conflict Data Program Georeferenced Event Dataset (UCDP GED) and Armed Conflict Location & Event Data (ACLED). UCDP GED provides georeferenced organized-violence records for 2014 to 2017, and ACLED provides detailed event and subevent descriptions for 2018 and 2019 (Raleigh et al., 2010; Sundberg & Melander, 2013). Event text was normalized and grouped into remote firepower, artillery or shelling, air or drone activity, mines or explosives, small arms, cyber or information activity, and civilian-directed violence. Figure 2 reports the annual source distribution and the frequency of these attack categories, while **Appendix C** records the complete extraction, cleaning, deduplication, and classification procedure.

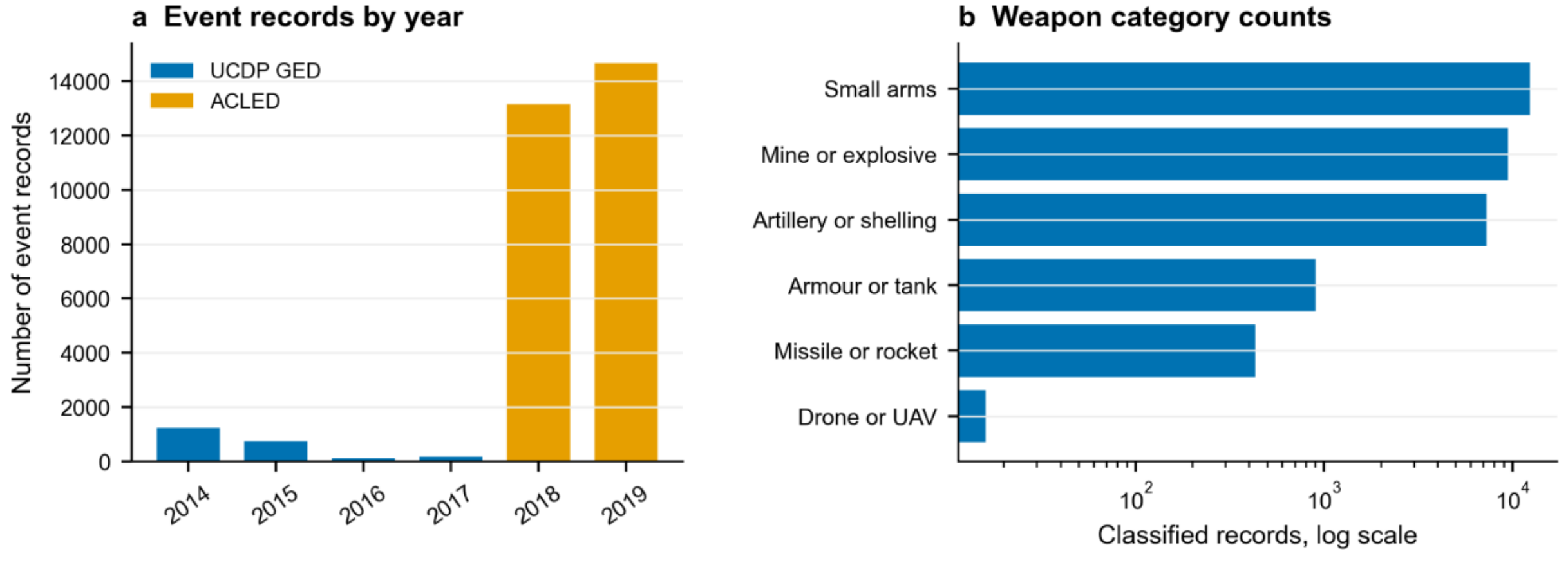

**Figure 2. Frequency distributions of Donbas conflict events and attack categories.** Both panels are bar distributions calculated from the 30,104 conflict records. Panel a reports source-specific event frequencies across the six conflict years. Panel b reports attack-category frequencies on a logarithmic horizontal scale so that common and rare categories remain visible.

Figure 2 shows that the recorded conflict environment is dominated by small arms, mines or explosives, and artillery or shelling, while air or drone and cyber or information events are less frequent. Annual counts are higher in the ACLED years because its event ontology and reporting density are more granular than the earlier UCDP GED coverage. The category distribution nevertheless provides a substantive map of the technologies and societal threats present in the Donbas record. This map motivates the later comparison between conflict channels and the semantic direction of Estonian research funding.

The external semantic corpus is drawn from OpenAlex, an open scholarly index that provides abstracts and research classifications (Priem et al., 2022). The operative paired file contains 5,449 war- or crisis-related abstracts and 5,449 nonwar abstracts matched by research field and publication year. Positive documents include explicit-war research and work on technologies or systems with plausible dual-use and crisis applications. **Appendix B** details the queries, inclusion rules, field-year matching, text preprocessing, and corpus distributions.

Field-by-year matching aligns the disciplinary and temporal composition of the positive and comparison corpora. The positive corpus is further organized around eight capability domains: engineering, materials, computer science, energy, medicine, life science, earth and environment, and agriculture and food systems. These domains reflect the technological clusters emphasized in defense-innovation, resilience, and crisis-response research, including digital systems, advanced materials, energy security, biotechnology, environmental sensing, and food-system resilience (European Commission, 2022; Mowery, 2012; North Atlantic Treaty Organization, 2020; Riebe et al., 2023). This organization supplies the external directions onto which ETIS project text is projected.

### 2.3 Semantic measurement

Available titles and abstracts are lowercased and represented with term-frequency inverse-document-frequency (TF-IDF) weights following Salton & Buckley (1988). The vocabulary contains unigrams and bigrams appearing in at least three documents and in no more than 85% of documents. English titles and abstracts are used when available, with original-language text retained as supplementary project information. **Appendix D** documents project-vector construction, annual aggregation, missing-text treatment, and the distribution of the resulting semantic outcomes.

$$w_{ij} = \left(1 + \ln f_{ij}\right) \ln \frac{N}{n_j} \tag{1}$$

In Equation (1), $f$ is the frequency of term $j$ in document $i$, $N$ is the number of documents, and $n$ is the number of documents containing term $j$. Each document vector is normalized to unit Euclidean length. The model is fitted jointly on the external corpus and the Estonia project text with a maximum of 60,000 features. This common representation makes project scores comparable across semantic dimensions.

For semantic dimension $k$, the centroid of positive documents is differenced from the centroid of matched comparison documents and normalized, following the contrastive logic of centroid-based relevance feedback (Rocchio, 1971). Separate axes are estimated for the general war corpus, explicit-war material, dual-use technology, and each of the eight capability domains. Dual-use refers to knowledge and technologies that can serve civilian purposes while also contributing to defense, security, or crisis operations, a concept developed in technology-policy scholarship and contemporary innovation governance (Alic, 1994; Brandt, 1994; Riebe et al., 2023). **Appendix E** develops the conceptual basis of the dual-use measure and maps it to the technological domains used here.

$$\mathbf{d}_k = \frac{\mathbf{p}_k - \mathbf{n}_k}{||\mathbf{p}_k - \mathbf{n}_k||_2} \tag{2}$$

In Equation (2), $p$ represents the positive-corpus centroid for semantic dimension $k$, and $n$ represents the matched-comparison centroid. Each project receives a signed projection onto direction $k$. Because the project vector and direction are normalized, Equation (3) is an inner product with the contrastive axis. Higher values indicate movement toward the positive rather than the comparison pole.

$$s_{ik} = \frac{\mathbf{g}_i{}^{\mathrm{T}}\mathbf{d}_k}{||\mathbf{g}_i||_2||\mathbf{d}_k||_2} \tag{3}$$

Four principal outcomes summarize project text. War all represents the general conflict axis, explicit war emphasizes direct conflict vocabulary, dual use represents the technical-application axis, and capability maximum is the largest raw score across the eight capability channels for each project. Annual means, medians, and top-decile shares capture movement at the center and upper tail of each distribution. The annual and project-level calculations are described in **Appendix D**, which also reports score distributions and aggregation checks.

**2.4 Funding and portfolio measures**

Financing-weighted semantic outcomes show whether records with larger reported amounts have different semantic scores from the typical record. Each score is weighted by its positive value in the ETIS FinancingInPeriodsTotal field and assigned to the recorded project year. The resulting series gives greater influence to projects with higher total financing recorded in ETIS. Equation (4) defines the financing-weighted score for dimension $k$ in year $t$.

$$S_{kt}^{w} = \frac{\sum_i^{N_t} A_{it}\, s_{ikt}}{\sum_i^{N_t} A_{it}} \tag{4}$$

Capability-channel balance is summarized from the eight annual channel scores. Nonnegative channel contributions are normalized to annual shares before Shannon entropy and the Herfindahl index are calculated. Higher entropy indicates a more even distribution of capability emphasis, while a higher Herfindahl value indicates greater concentration. These measures describe whether semantic reorientation is broadly distributed across capabilities or concentrated in a narrower subset.

$$H_t = -\sum_j^J q_{jt}\ln q_{jt}, \mathrm{HHI}_t = \sum_j^J q_{jt}^2 \quad (5)$$

### 2.5 Empirical specifications and inference

The main annual specification is a segmented trend model centered on 2014. It separates the pre-2014 trend, a level difference beginning in 2014, and a change in slope during 2014 to 2019. The same structure is estimated for unweighted and financing-weighted annual outcomes. This specification provides a transparent summary of whether the direction and pace of the annual semantic series changed at the Donbas-war breakpoint.

$$y_t = \alpha + \beta_1(t - 2000) + \beta_2\mathrm{Post}_{2014,t} + \beta_3(t\text{-}2014)^+ + \varepsilon_t \quad (6)$$

Coefficient $\beta_2$ captures the post-2014 level difference relative to the projected pre-2014 trend, and $\beta_3$ captures the slope change. Inference uses heteroskedasticity- and autocorrelation-consistent (HAC) standard errors (SEs) with one lag (Newey & West, 1987). Project-level annual effects additionally adjust for abstract availability, English-text availability, text length and its square, and descriptive field composition. These controls align annual comparisons with the observable composition of the ETIS record stream.

Exploratory conflict-channel alignment is evaluated with Spearman correlations for 2014 to 2019, using contemporaneous and one-year-lagged semantic outcomes. Complete enumeration supplies two-sided permutation probabilities for the six-year sequence (Ernst, 2004). The full set of event-channel coefficients and exact probabilities is reported in **Appendix F**. The analysis emphasizes the sign, timing, and substantive correspondence between recorded conflict technologies and the research-funding measures.

## 3 Semantic change in Estonia's recorded research portfolio

### 3.1 Semantic measurement and adjusted annual patterns

The operative benchmark contains 5,449 war- or crisis-related abstracts and 5,449 field-by-year matched comparison abstracts. Each ETIS research-funding record receives signed projections for the general-war, explicit-war, dual-use, and eight capability axes. Table 2 defines the analytical samples and the interpretation of the four principal indicators. The definitions establish how positive values represent movement toward the vocabulary characteristic of the relevant reference corpus.

**Table 2. Semantic samples and indicator definitions.**

| Component | Sample size | Role | Interpretation |
|---|---|---|---|
| War/crisis corpus | 5,449 | Positive contrastive poles | Retrospective reference text |
| Matched comparison corpus | 5,449 | Field-year comparison poles | Controls reference-corpus composition |
| ETIS target portfolio | 17,952 | Projected project records | Mixed-source Estonian research portfolio |
| Upper-tail measure | Annual ETIS records | Share above fixed pre-2014 P90 | Movement into baseline upper tail |

Note: TF-IDF indicators use the same operative 5,449-by-5,449 paired corpus and vocabulary. Higher values denote projections toward the positive reference centroid. Annual means, medians, and top-decile shares summarize different locations in the project distribution.

As shown in Table 2, the matched OpenAlex corpus provides the semantic contrast, while ETIS supplies the research-funding portfolio to which that contrast is applied. The general-war score captures broad crisis proximity, the explicit-war score isolates overt conflict language, and the dual-use score captures technical applications spanning civilian and security domains. Capability maximum records the strongest of eight domain-specific alignments for each project. Figure 3 then traces adjusted annual change in explicit-war, dual-use, and maximum-capability projections relative to 2013.

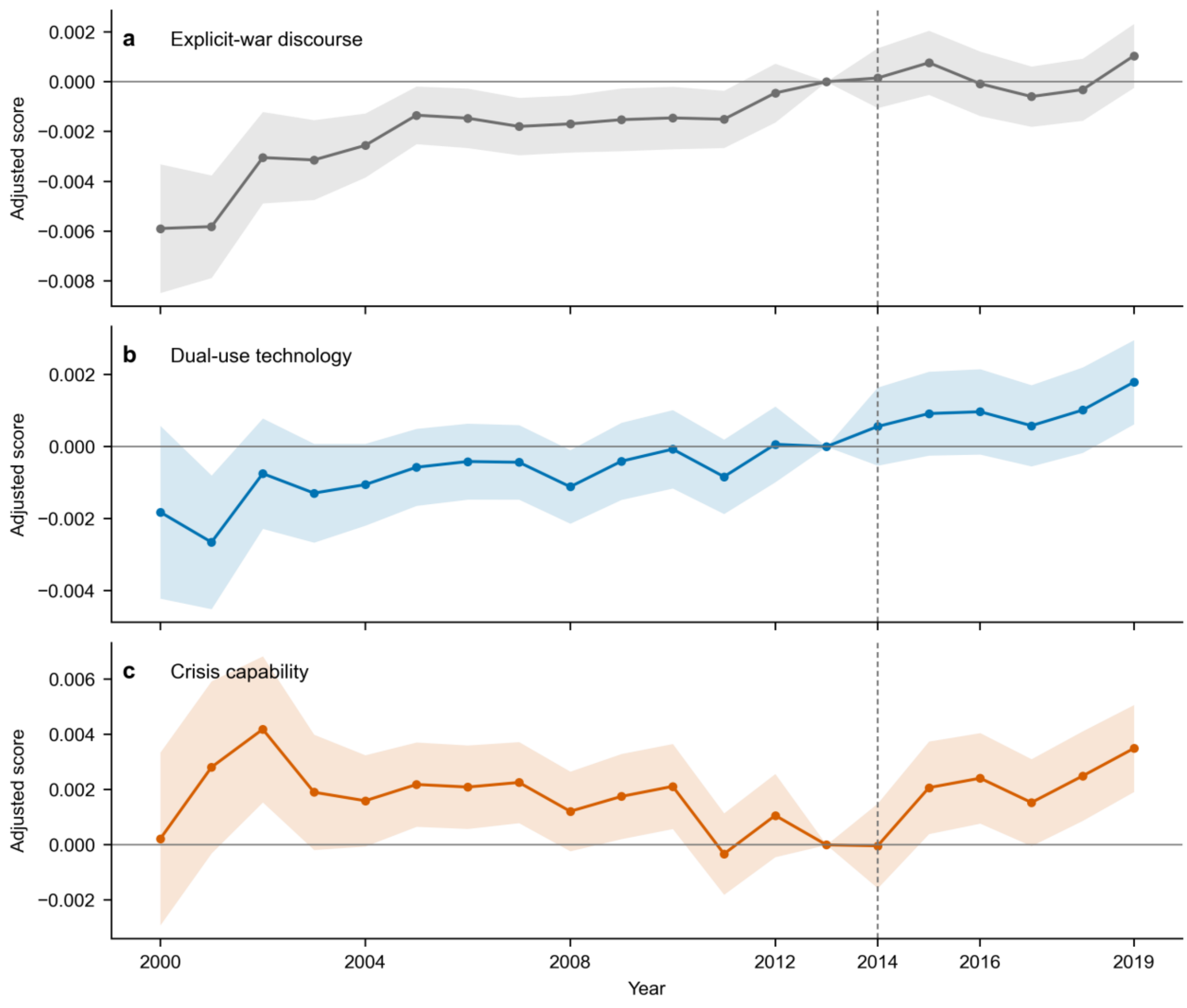

**Figure 3. Adjusted annual projections on explicit-war, dual-use, and capability axes.**

Figure 3 reveals a clear separation between overt conflict language and capability-oriented research language. Explicit-war projections do not rise after 2014, whereas dual-use and maximum-capability projections remain generally above their 2013 reference during 2014 to 2019. The contrast is strongest in the later post-breakpoint years, indicating that the portfolio moved toward technological preparedness more than toward direct war terminology. This pattern is consistent with mission-oriented adaptation that builds on established scientific capabilities rather than relabeling projects in military terms (Hill et al., 2025; Mowery, 2012).

### 3.2 Segmented changes in semantic projections

The principal segmented estimates show that post-2014 change is concentrated in dual-use and maximum-capability language. As shown in Table 3, the strongest coefficients appear in slope changes for the mean, median, and upper tail rather than in immediate level shifts. This distributional agreement indicates that the pattern is visible both in the typical project and among projects with especially strong semantic alignment. Figure 4 subsequently compares the post-2014 slope coefficients and their confidence intervals across the core outcomes.

**Table 3. Segmented-trend coefficients for semantic outcomes.**

| Indicator | Post 2014 level | SE | Post 2014 slope | SE |
|---|---|---|---|---|
| War related top ten percentile share | 0.0072 | 0.0090 | 0.0056** | 0.0022 |
| War related mean projection | 0.0005 | 0.0005 | 0.0000 | 0.0002 |
| Explicit war mean projection | -0.0007 | 0.0007 | -0.0003* | 0.0002 |
| Dual use mean projection | 0.0014* | 0.0009 | 0.0003** | 0.0002 |
| Dual use median projection | 0.0017 | 0.0011 | 0.0006*** | 0.0002 |
| Dual use top ten percentile share | 0.0078 | 0.0101 | 0.0060*** | 0.0022 |
| Capability max mean projection | 0.0029 | 0.0019 | 0.0014*** | 0.0003 |
| Capability max median projection | 0.0032 | 0.0020 | 0.0018*** | 0.0003 |
| Capability max top ten percentile share | 0.0068 | 0.0134 | 0.0084*** | 0.0024 |

Note: $N = 20$ annual observations from 2000 to 2019. Estimates use HAC standard errors with one lag. Two-sided significance levels are * $p < 0.1000$, ** $p < 0.0500$, and *** $p < 0.01$. Coefficients describe a 2014 level difference and post-2014 slope change relative to the pre-2014 trend.

The dual-use mean slope rises by 0.0003 per year (SE = 0.0002, $p = 0.0460$), while the median and upper-tail slopes rise by 0.0006 (SE = 0.0002, $p = 0.0070$) and 0.0060 (SE = 0.0022, $p = 0.0070$). Maximum-capability effects are larger and more precise, with slope changes of 0.0014 for the mean (SE = 0.0003, $p < 0.0010$), 0.0018 for the median (SE = 0.0003, $p < 0.0010$), and 0.0084 for the upper tail (SE = 0.0024, $p = 0.0010$). The explicit-war mean instead has a negative slope of -0.0003 (SE = 0.0002, $p = 0.0700$), which separates technical preparedness from overt conflict vocabulary. These estimates support the view that mission pressures often enter research systems through adjacent technical capabilities and problem-oriented applications (Foray et al., 2012; Mowery, 2012).

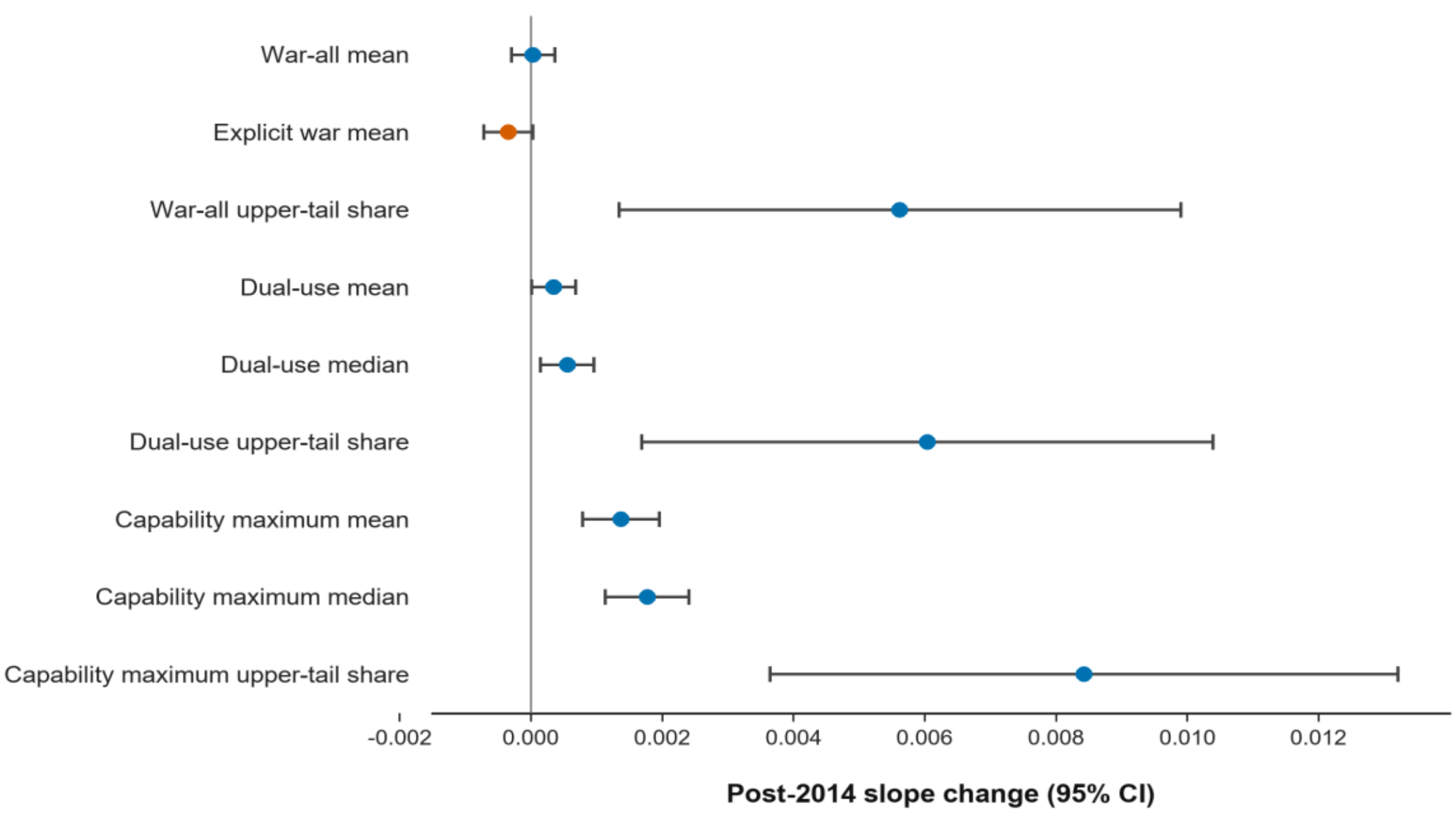

**Figure 4. Post-2014 slope changes in core semantic outcomes.**

Figure 4 makes the coefficient pattern visually explicit. Positive estimates cluster in the dual-use and maximum-capability measures, and their intervals are predominantly separated from zero, while the explicit-war mean points downward. The largest standardized separation occurs for maximum-capability mean and median scores, showing that the change extends beyond a small number of extreme projects. The figure therefore reinforces the interpretation that the post-2014 portfolio became more aligned with crisis-relevant capabilities without a corresponding rise in explicit-war framing.

### 3.3 Conflict-record coverage and exploratory alignment

The conflict file provides a detailed account of the technologies and forms of violence present in the Donbas theatre. Table 4 reports annual source coverage together with the attack categories extracted from event descriptions. The table shows the transition from UCDP GED coverage in 2014 to 2017 to the more granular ACLED stream in 2018 and 2019. These annual distributions establish the empirical basis for the exploratory alignment shown in Figure 5.

**Table 4. Annual conflict-record coverage by source.**

| Year | Source | Records | Fatalities | Remote-firepower records | Civilian-violence records |
|---|---|---|---|---|---|
| 2014 | UCDP GED | 1238 | 3821 | 43 | 0 |
| 2015 | UCDP GED | 741 | 1365 | 42 | 0 |
| 2016 | UCDP GED | 126 | 253 | 36 | 0 |
| 2017 | UCDP GED | 184 | 399 | 78 | 0 |
| 2018 | ACLED | 13157 | 861 | 3793 | 21 |
| 2019 | ACLED | 14658 | 355 | 3461 | 24 |

Note: Counts describe records in the merged Donbas conflict file. UCDP GED supplies 2014 to 2017 and ACLED supplies 2018 to 2019. Category counts are derived from standardized event descriptions and classification rules reported in **Appendix C**.

As shown in Table 4, recorded conflict activity rises sharply in 2018 and 2019, with remote firepower becoming especially prominent in the final year. Artillery, explosives, and small-arms categories remain central across the conflict record, while cyber and drone-related events occupy smaller but strategically distinctive segments. Figure 5 places these channels beside contemporaneous and lagged Estonian semantic outcomes. The comparison is designed to identify whether changes in the content of the conflict environment are mirrored by changes in research-funding language.

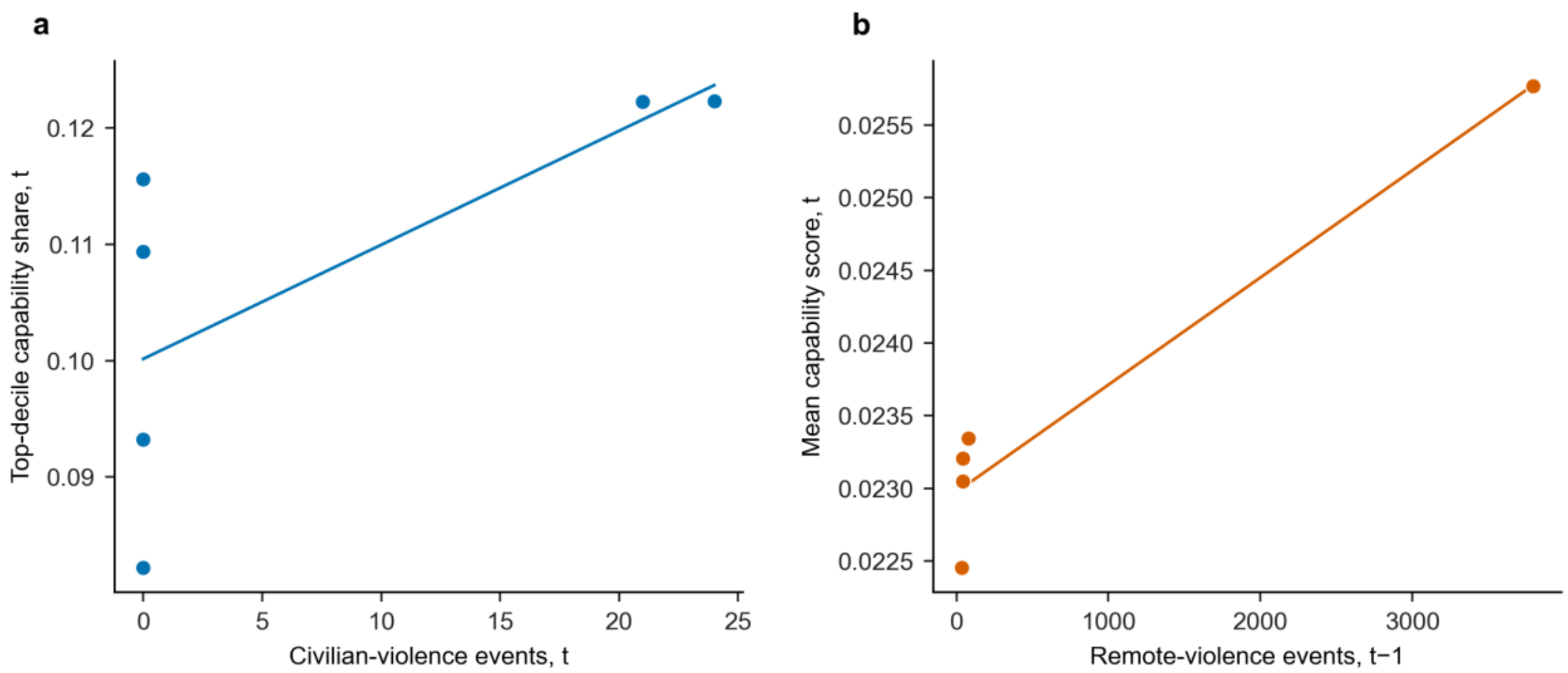

**Figure 5. Recorded conflict channels and semantic outcomes during 2014 to 2019.**

Figure 5 shows that the strongest event-semantic alignment occurs between total recorded conflict activity and the following year's dual-use mean. Across the five available lagged years, the two series have identical rank ordering, with Spearman $\rho = 1.0000$ and an exact two-sided $p$ value of 0.0250. This coefficient describes rank correspondence in a short annual sequence rather than the magnitude of a linear effect. The remaining coefficients vary in sign and size, indicating heterogeneous correspondence across conflict technologies and societal threats.

### 3.4 Award-size weighting of semantic reorientation

Financing-weighted scores reveal whether semantic orientation differs when larger recorded projects receive greater influence. Table 5 compares project-weighted and financing-weighted level and slope coefficients across the main semantic axes. Figure 6 then traces the adjusted dual-use paths under the two weighting rules. Together, the table and figure locate the post-2014 signal within the award-size distribution.

**Table 5. Unweighted and financing-weighted segmented estimates.**

| Outcome | Post-2014 level | SE | Post-2014 slope | SE |
|---|---|---|---|---|
| War-all mean | 0.0005 | 0.0005 | 0.0000 | 0.0002 |
| War-all financing-weighted | 0.0029** | 0.0013 | -0.0001 | 0.0002 |
| Explicit-war mean | -0.0007 | 0.0007 | -0.0003* | 0.0002 |
| Explicit-war financing-weighted | 0.0021* | 0.0012 | -0.0003 | 0.0002 |
| Dual-use mean | 0.0014* | 0.0009 | 0.0003** | 0.0002 |
| Dual-use financing-weighted | 0.0031** | 0.0014 | -0.0000 | 0.0002 |
| Capability maximum mean | 0.0029 | 0.0019 | 0.0014*** | 0.0003 |
| Capability maximum financing-weighted | 0.0057** | 0.0027 | 0.0002 | 0.0003 |

Note: $N$ = 20 annual observations. Financing weights use positive values of ETIS FinancingInPeriodsTotal, a total project-financing field assigned to the recorded project year. Two-sided significance levels are * $p < 0.1000$, ** $p < 0.0500$, and *** $p < 0.01$.

As shown in Table 5, financing-weighted war-all, dual-use, and maximum-capability scores have positive post-2014 level differences of 0.0029 ($p = 0.0250$), 0.0031 ($p = 0.0300$), and 0.0057 ($p = 0.0330$). Their subsequent slope changes are small, whereas the unweighted dual-use series has a positive slope change of 0.0003 ($p = 0.0460$). This contrast indicates that larger projects moved first through an upward level difference, while gradual change is more visible across the broader project population. Similar differences between programme scale and portfolio-wide diffusion are consistent with evidence that funding design shapes where public R&D effects enter an innovation system (Azoulay et al., 2019; Howell et al., 2025).

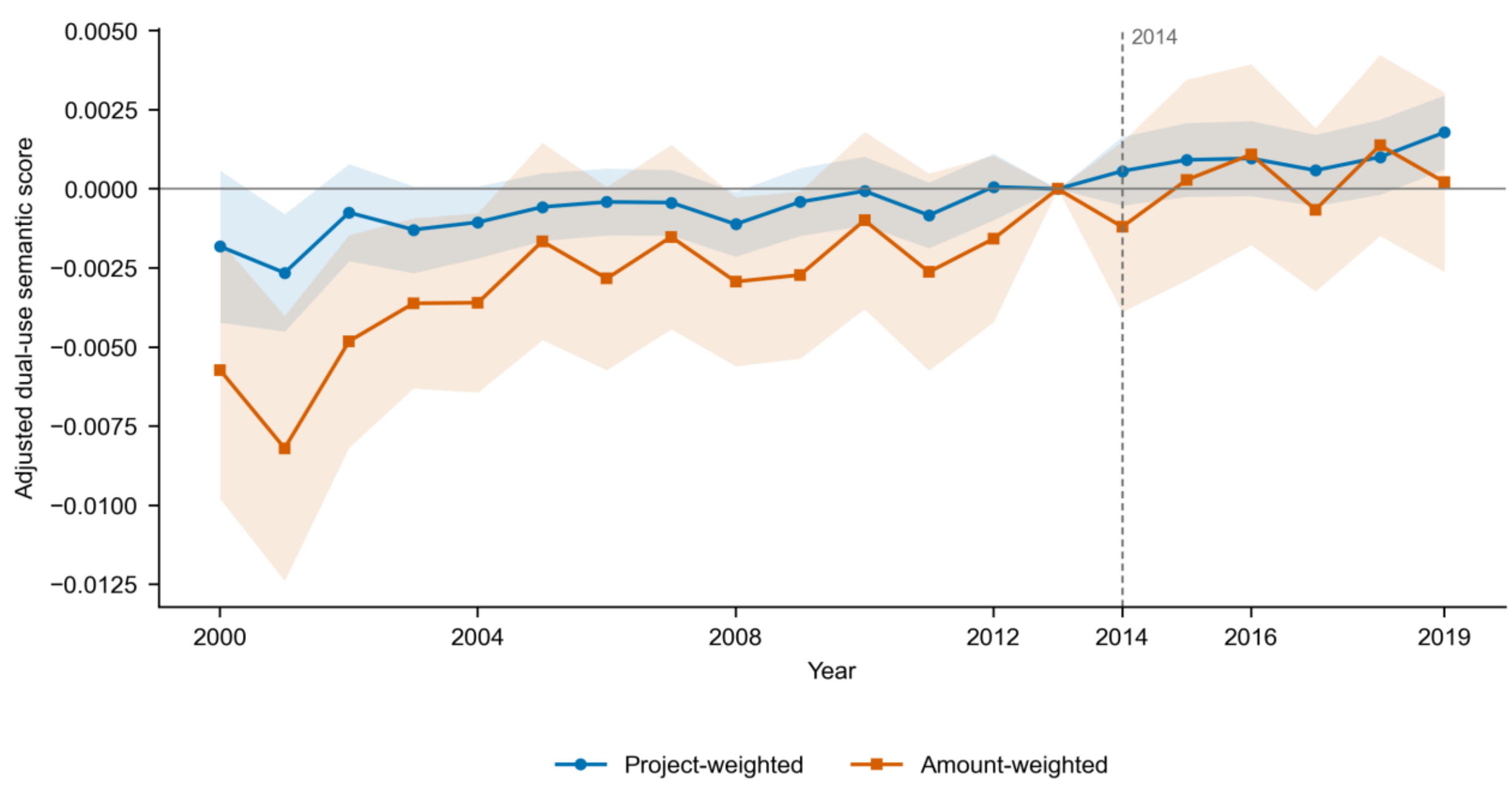

**Figure 6. Adjusted project-weighted and financing-weighted dual-use projections.**

Figure 6 shows that the financing-weighted dual-use series is more variable than the project-weighted series and moves above its 2013 reference in several post-2014 years. The project-weighted path is smoother and exhibits the gradual rise summarized by the segmented slope coefficient. The combination of an early weighted level shift and a later unweighted trend indicates that capability-oriented language was initially more pronounced among larger awards before becoming more broadly distributed. The two weighting rules therefore locate the semantic shift in different portions of the award distribution.

### 3.5 Capability-channel heterogeneity and balance

The final empirical block separates the eight semantic capability channels. Table 6 compares pre-2014 and post-2014 means and reports the post-breakpoint trend estimate for each channel. Figure 7 places these channel differences beside annual entropy, revealing how the balance of capability attention changed over time. The combined evidence identifies both the substantive domains associated with reorientation and the degree to which the portfolio concentrated around them.

**Table 6. Capability-channel scores before and after 2014.**

| Channel | Pre 2014 mean | Post 2014 mean | Simple change | Post 2014 trend | Post p |
|---|---|---|---|---|---|
| Computer science | 0.0072 | 0.0091 | 0.0019 | -0.0000 | 0.7540 |
| Energy | 0.0076 | 0.0093 | 0.0017 | -0.0002 | 0.2670 |
| Engineering | 0.0096 | 0.0111 | 0.0015 | -0.0002 | 0.2270 |
| Agriculture food | 0.0017 | 0.0026 | 0.0009 | -0.0002 | 0.1080 |
| Medicine | 0.0041 | 0.0050 | 0.0008 | 0.0001 | 0.4520 |
| Earth environment | 0.0057 | 0.0061 | 0.0004 | -0.0002 | 0.1140 |
| Life sciences | 0.0055 | 0.0044 | -0.0012 | -0.0002 | 0.1860 |
| Materials | 0.0040 | 0.0025 | -0.0014 | -0.0004** | 0.0260 |

Note: Rows are the eight semantic capability channels. Simple changes compare period means, and post-2014 trend $p$ values come from the segmented model. Positive differences indicate stronger average semantic alignment after 2014.

As shown in Table 6, computer science records the largest positive period difference at 0.0019, followed by energy at 0.0017, engineering at 0.0015, agriculture and food at 0.0009, and medicine at 0.0008. Materials and life science move in the opposite direction, and the materials post-2014 trend is -0.0004 with $p$ = 0.026. The cross-channel pattern therefore combines stronger digital, energy, engineering, food-system, and medical orientation with weaker emphasis in selected upstream scientific domains. This configuration corresponds closely to the capability clusters highlighted in NATO technology assessments and European analyses of the war's effects on energy, supply chains, security, and research demand (European Commission, 2022; North Atlantic Treaty Organization, 2020).

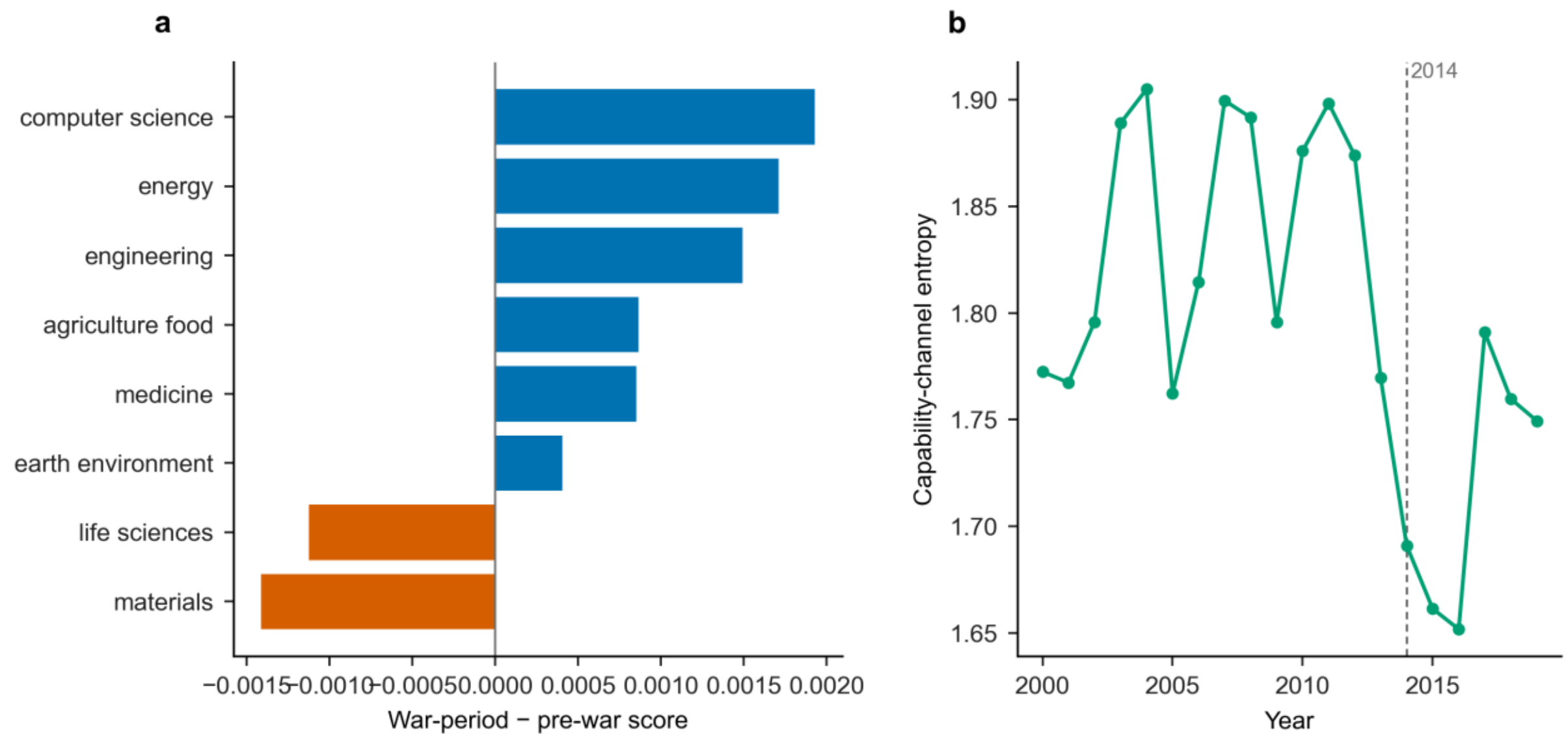

**Figure 7. Capability-channel mean differences and annual channel entropy.**

Figure 7 shows that the capability profile became temporarily more concentrated after 2014. Entropy falls around the breakpoint, reaches its lowest value in 2016, recovers in 2017, and remains below most pre-2014 observations in 2018 and 2019. The left panel shows that this narrowing coincides with positive mean differences in computer science, energy, engineering, agriculture and food, and medicine, rather than an equal rise across all channels. The pattern suggests a selective preparedness response centered on a limited set of operationally relevant capabilities, consistent with mission-oriented policy's tendency to coordinate research around prioritized societal problems (Foray et al., 2012; Kattel & Mazzucato, 2018).

# 4 Discussion and conclusion

This study combines 17,952 ETIS research-funding records, a matched OpenAlex semantic corpus, and geocoded Donbas conflict events to examine how a regional-security shock is reflected in the direction of a neighboring country's research portfolio. Project texts are projected onto explicit-war, dual-use, and eight capability directions, then summarized through distributional, financing-weighted, and segmented annual measures. The empirical results identify where semantic change occurred, how it was distributed across project size, and which capability domains contributed most strongly. The analysis thereby connects geopolitical change to the evolving content of funded research in Estonia.

The central implication is that geopolitical exposure can enter research policy through the reinterpretation of existing scientific capabilities. Estonia's portfolio does not need to adopt the vocabulary of war for its technical agenda to become more relevant to preparedness. Research directions in computation, energy, engineering, health, and supply resilience already have civilian foundations, yet their strategic value rises when the regional security environment changes. This mechanism connects threat perception to research priorities through capability development rather than administrative relabeling.

Mission-oriented research policy clarifies why technical language is a sensitive indicator of strategic adjustment. Defense and crisis missions often draw on technologies with substantial civilian applications, so the relevant policy boundary runs through uses and capabilities rather than conventional sector labels (Alic, 1994; Brandt, 1994; Mowery, 2012). In this setting, dual-use orientation is a bridge between an established knowledge base and a changing set of public problems. The Estonian case adds a regional-security dimension to the literature on how missions reorganize the direction of innovation systems.

The difference between project weighting and financing weighting also has an institutional interpretation. Larger awards usually involve more coordination, infrastructure, and organizational commitment than smaller projects, making their thematic orientation especially informative about portfolio direction. The subsequent appearance of related language across the unweighted series points to a broader reorganization of project content rather than an isolated movement among a few records. Award size and portfolio breadth thus reveal complementary dimensions of how strategic priorities become embedded in a research system.

The mix of capability channels shows that strategic adaptation is selective. Digital systems, energy, engineering, food systems, and medicine connect immediate preparedness needs to Estonia's existing technological and administrative strengths. Materials and life science follow different trajectories, which suggests that geopolitical relevance is filtered through national specialization rather than imposed uniformly across science. NATO technology assessments and European policy analyses similarly organize preparedness around interacting digital, infrastructural, energy, health, and supply capabilities (European Commission, 2022; North Atlantic Treaty Organization, 2020).

Changes in channel balance reveal a second stage of adaptation. Concentration can help a research system coordinate attention when a new strategic problem first becomes salient, while later broadening allows those priorities to connect with a wider knowledge base. The observed sequence fits this movement from focused attention to partial recombination across capabilities. This interpretation is consistent with mission-oriented policy accounts that combine directionality with distributed problem-solving capacity (Foray et al., 2012; Kattel & Mazzucato, 2018).

Estonia's institutional history gives this mechanism a specific national form. Cyber-security institutions, NATO and European integration, and experience in linking digital infrastructure to national resilience created channels through which new security concerns could be translated into research problems (Crandall & Allan, 2015; Wrange & Bengtsson, 2019). The Research and Development, Innovation and Entrepreneurship Strategy 2021 to 2035 later connected knowledge, digital transformation, energy, and resilience within a common development framework. Strategic adaptation therefore appears as an extension and recombination of institutional capacity rather than a departure from Estonia's prior research trajectory.

The methodological implication extends beyond this case. Administrative fields locate projects within formal disciplines, whereas contrastive semantic directions reveal how the meaning of technically similar work changes in relation to an external problem environment. Combining the center, upper tail, and financing-weighted portions of the distribution further shows whether reorientation is diffuse or institutionally concentrated. This framework offers a comparative strategy for studying how neighboring research systems absorb geopolitical change through project content.

Estonia remains a single national case, and the conflict record changes source after 2017. The empirical design therefore identifies longitudinal associations rather than an average causal effect. This boundary limits how far the estimated pattern can be generalized beyond the Estonian setting. Another limitation is that the ETIS records do not capture non-public or classified defence research funding. The estimates therefore describe semantic shifts in publicly recorded research funding and may understate changes occurring in restricted defence-related portfolios.

## References

Alic, J. A. (1994). The dual use of technology: Concepts and policies. *Technology in Society*, *16*(2), 155–172. https://doi.org/10.1016/0160-791X(94)90027-2

Azoulay, P., Graff Zivin, J. S., Li, D., & Sampat, B. N. (2019). Public R&D investments and private-sector patenting: Evidence from US National Institutes of Health (NIH) funding rules. *The Review of Economic Studies*, *86*(1), 117–152. https://doi.org/10.1093/restud/rdy034

Brandt, L. (1994). Defense conversion and dual-use technology. *Policy Studies Journal*, *22*(2), 359–370. https://doi.org/10.1111/j.1541-0072.1994.tb01474.x

Caldara, D., & Iacoviello, M. (2022). Measuring geopolitical risk. *American Economic Review*, *112*(4), 1194–1225. https://doi.org/10.1257/aer.20191823

Crandall, M. (2014). Soft security threats and small states: The case of Estonia. *Defence Studies*, *14*(1), 30–55. https://doi.org/10.1080/14702436.2014.890334

Crandall, M., & Allan, C. (2015). Small states and big ideas: Estonia's battle for cyber-security norms. *Contemporary Security Policy*, *36*(2), 346–368. https://doi.org/10.1080/13523260.2015.1061765

Ernst, M. D. (2004). Permutation methods: A basis for exact inference. *Statistical Science*, *19*(4), 676–685. https://doi.org/10.1214/088342304000000396

European Commission. (2022). *EU research and innovation and the invasion of Ukraine: Main channels of impact*. https://research-and-innovation.ec.europa.eu/knowledge-publications-tools-and-data/publications/all-publications/eu-research-and-innovation-and-invasion-ukraine-main-channels-impact_en

Foray, D., Mowery, D. C., & Nelson, R. R. (2012). Public R&D and social challenges: What lessons from mission R&D programs. *Research Policy*, *41*(10), 1697–1702. https://doi.org/10.1016/j.respol.2012.07.011

Ganguli, I. (2017). Saving Soviet science: The impact of grants when government R&D funding disappears. *American Economic Journal: Applied Economics*, *9*(2), 165–201. https://doi.org/10.1257/app.20160180

Gross, D. P., & Sampat, B. N. (2023). America, jump-started: World War II R&D and the takeoff of the US innovation system. *American Economic Review*, *113*(12), 3323–3356. https://doi.org/10.1257/aer.20221365

Hill, R., Yin, Y., Stein, C., Wang, X., Wang, D., & Jones, B. F. (2025). The pivot penalty in research. *Nature*, *642*, 999–1006. https://doi.org/10.1038/s41586-025-09048-1

Howell, S. T., Rathje, J., Van Reenen, J., & Wong, J. (2025). Opening up military innovation: Causal effects of reforms to US defense research. *Journal of Political Economy*, *133*(11), 3605–3651. https://doi.org/10.1086/737235

Kattel, R., & Mazzucato, M. (2018). Mission-oriented innovation policy and dynamic capabilities in the public sector. *Industrial and Corporate Change*, *27*(5), 787–801. https://doi.org/10.1093/icc/dty032

Kuzhabekova, A. (2024). Regional spillover effect of 2022 sanctions against Russia on scholarly publications. *Learned Publishing*, *37*(4), e1619. https://doi.org/10.1002/leap.1619

Mazzucato, M. (2018). Mission-oriented innovation policies: Challenges and opportunities. *Industrial and Corporate Change*, *27*(5), 803–815. https://doi.org/10.1093/icc/dty034

Moretti, E., Steinwender, C., & Van Reenen, J. (2025). The intellectual spoils of war. *Defense R&D, Productivity, and International Spillovers. The Review of Economics and Statistics*, *107*(1), 14–27. https://doi.org/10.1162/rest_a_01293

Mowery, D. C. (2012). Defense-related R&D as a model for grand challenges technology policies. *Research Policy*, *41*(10), 1703–1715. https://doi.org/10.1016/j.respol.2012.03.027

Newey, W. K., & West, K. D. (1987). A simple, positive semi-definite, heteroskedasticity and autocorrelation consistent covariance matrix. *Econometrica*, *55*(3), 703–708. https://doi.org/10.2307/1913610

North Atlantic Treaty Organization. (2020). *Science & Technology Trends 2020–2040: Exploring the S&T Edge*. https://www.nato.int/nato_static_fl2014/assets/pdf/2020/4/pdf/190422-ST_Tech_Trends_Report_2020-2040.pdf

Priem, J., Piwowar, H., & Orr, R. (2022). OpenAlex: A fully-open index of scholarly works, authors, venues, institutions, and concepts. *Quantitative Science Studies*, *3*(3), 687–708. https://doi.org/10.1162/qss_a_00197

Raleigh, C., Linke, A., Hegre, H., & Karlsen, J. (2010). Introducing ACLED: An armed conflict location and event dataset. *Journal of Peace Research*, *47*(5), 651–660. https://doi.org/10.1177/0022343310378914

Riebe, T., Schmid, S., & Reuter, C. (2023). Dual-use and trustworthy. *A Mixed Methods Analysis of AI Diffusion between Civilian and Defense R&D. Science and Engineering Ethics*, *29*. https://doi.org/10.1007/s11948-023-00431-1

Rocchio, J. J. (1971). Relevance feedback in information retrieval. In G. Salton (Ed.), *The SMART Retrieval System: Experiments in Automatic Document Processing* (pp. 313–323). Prentice-Hall.

Salton, G., & Buckley, C. (1988). Term-weighting approaches in automatic text retrieval. *Information Processing & Management*, *24*(5), 513–523. https://doi.org/10.1016/0306-4573(88)90021-0

Schot, J., & Steinmueller, W. E. (2018). Three frames for innovation policy: R&D, systems of innovation and transformative change. *Research Policy*, *47*(9), 1554–1567. https://doi.org/10.1016/j.respol.2018.08.011

Sundberg, R., & Melander, E. (2013). Introducing the UCDP Georeferenced Event Dataset. *Journal of Peace Research*, *50*(4), 523–532. https://doi.org/10.1177/0022343313484347

Waldinger, F. (2016). Bombs, brains, and science: The role of human and physical capital for the creation of scientific knowledge. *The Review of Economics and Statistics*, *98*(5), 811–831. https://doi.org/10.1162/REST_a_00565

Wrange, J., & Bengtsson, R. (2019). Internal and external perceptions of small state security: The case of Estonia. *European Security*, *28*(4), 449–472. https://doi.org/10.1080/09662839.2019.1665517

## Appendix A. ETIS research-funding records and descriptive mapping

This appendix documents the construction of the 17,952-record Estonian Research Information System (ETIS) analytical file. Records were retained when a project year between 2000 and 2019 could be assigned from the recorded start date or project-period metadata. Exact identifiers were used to control duplicates, text fields were standardized without altering substantive content, and financing was converted to numeric positive values where available. Table A1 summarizes the principal completeness indicators used in cleaning and adjustment. The seven descriptive fields were assigned from ETIS labels first and text anchors second, providing the coverage summary used in Figure 1.

**Table A1. Completeness of the ETIS analytical file.**

| Indicator | Value |
|---|---|
| Records | 17,952 |
| Pre-2014 records | 11,555 |
| 2014-2019 records | 6,397 |
| Original abstract | 90.3% |
| English abstract | 59.1% |
| Positive financing | 97.9% |

Figure A1 extends the aggregate rates in Table A1 by showing how original-abstract, English-abstract, and positive-financing coverage varies across project years. Figure A1 shows financing coverage above 95% in every year. Original-abstract coverage remains high through 2015 and declines thereafter. English-abstract coverage is lower, but it is comparatively stable around the breakpoint, remaining at approximately 53% to 64% from 2008 to 2018 without an abrupt change in 2014. The main analysis additionally controls for abstract and English-text availability.

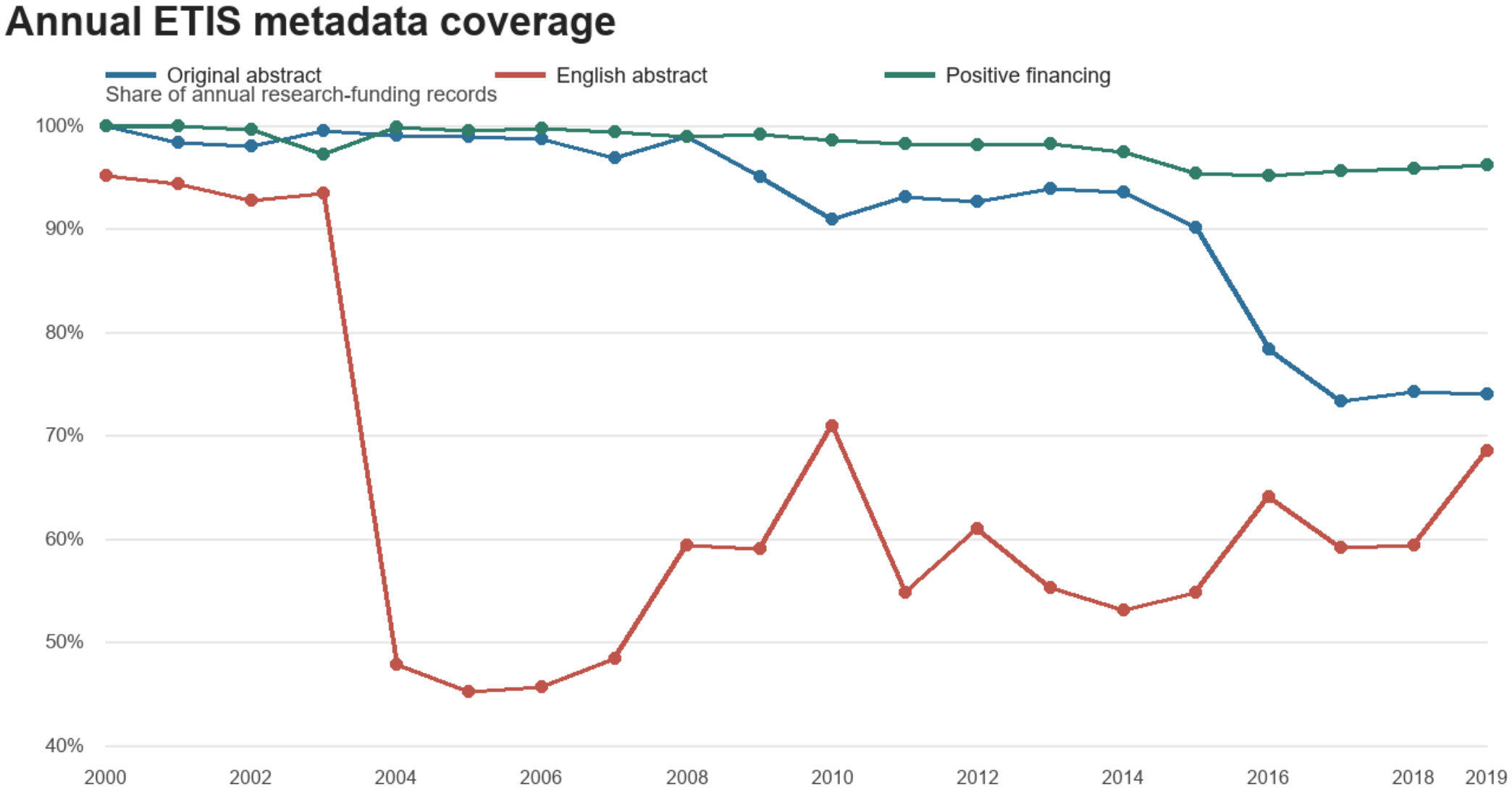

**Figure A1. Annual metadata coverage in the ETIS analytical file.**

# Appendix B. War and matched nonwar reference corpora

OpenAlex abstracts were retrieved for 2000 to 2024 using concept, title, and abstract criteria that identify armed conflict, security crises, and crisis-relevant technologies (Priem et al., 2022). Records without reconstructable abstracts were removed, exact work identifiers were deduplicated, and text was normalized to lowercase Unicode before vectorization. Each positive document was paired with a nonwar document in the same publication year and broad research field. Matching preserved 5,449 documents in each group and created a balanced contrast for centroid construction.

Table B1 establishes the one-to-one field-by-year matched design. Figure B1 complements Table B1 by showing how the positive corpus is divided between explicit-war and technical-application material and how its documents are distributed across the leading research fields.

**Table B1. OpenAlex reference-corpus design.**

| Corpus group | Documents | Matching condition |
|---|---|---|
| War/crisis positive | 5,449 | Field and year |
| Nonwar comparison | 5,449 | Field and year |

Figure B1 shows that technical-application abstracts form the larger semantic tier, with 3,211 documents compared with 2,238 explicit-war documents. Social sciences constitute the largest field group, followed by engineering, medicine, psychology, health professions, and environmental science. The extended field distribution allows the semantic direction to incorporate both societal-threat research and technical capability research. Exact field-year matching reproduces the same disciplinary and temporal composition in the nonwar comparison corpus.

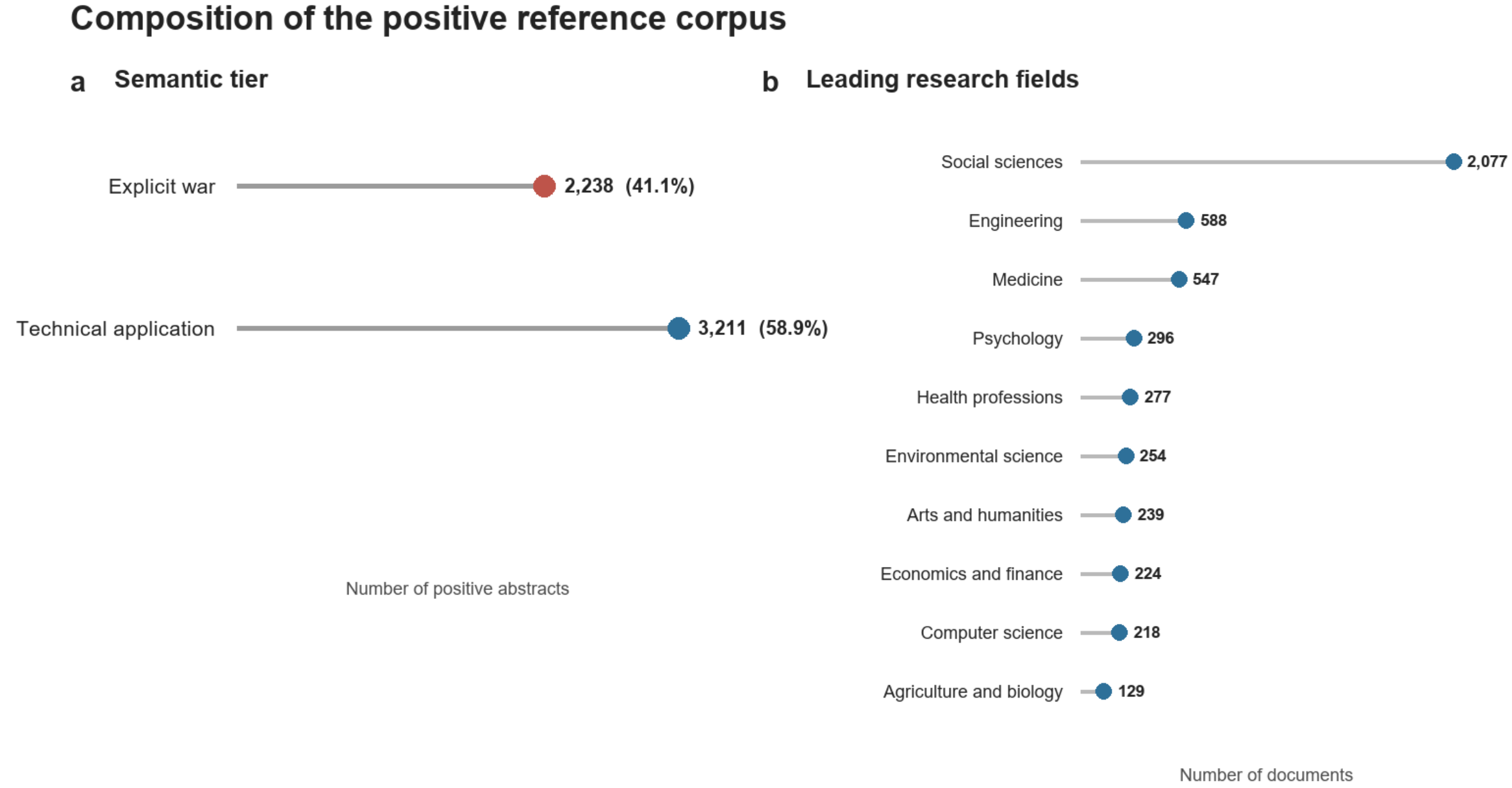

**Figure B1. Composition of the positive OpenAlex corpus by semantic tier and research field.**

# Appendix C. Donbas conflict-event construction

The Donbas file combines the Uppsala Conflict Data Program Georeferenced Event Dataset (UCDP GED) for 2014 to 2017 with Armed Conflict Location & Event Data (ACLED) records for 2018 and 2019 (Raleigh et al., 2010; Sundberg & Melander, 2013). Coordinates and administrative-location fields were used to retain observations within the study region, event dates were parsed to calendar years, and repeated source identifiers were removed. Normalized event descriptions were classified with transparent multilingual term rules covering small arms, mines or improvised explosive devices, artillery or shelling, armour or tanks, missiles or rockets, and drones or unmanned aerial vehicles. The remote-violence measure used in the event-semantic analysis combines missiles or rockets, artillery or shelling, and drones or unmanned aerial vehicles, while civilian-directed violence is identified from the standardized event-type field.

Table C1 reports the annual number of records in the final conflict-event file. Figure C1 adds information not contained in Table C1 by comparing attack-category totals and displaying each category's share of annual records from 2014 to 2019.

**Table C1. Annual distribution of Donbas conflict records.**

| Year | Records |
|---|---|
| 2014 | 1,238 |
| 2015 | 741 |
| 2016 | 126 |
| 2017 | 184 |
| 2018 | 13,157 |
| 2019 | 14,658 |

Figure C1 shows that small arms, mines or explosives, and artillery or shelling form the largest classified groups. Armour or tanks and missiles or rockets are less frequent, while drones form a smaller distinct category. Category shares vary markedly across years rather than following the annual total in Table C1. These patterns define the event context used in the subsequent semantic comparisons.

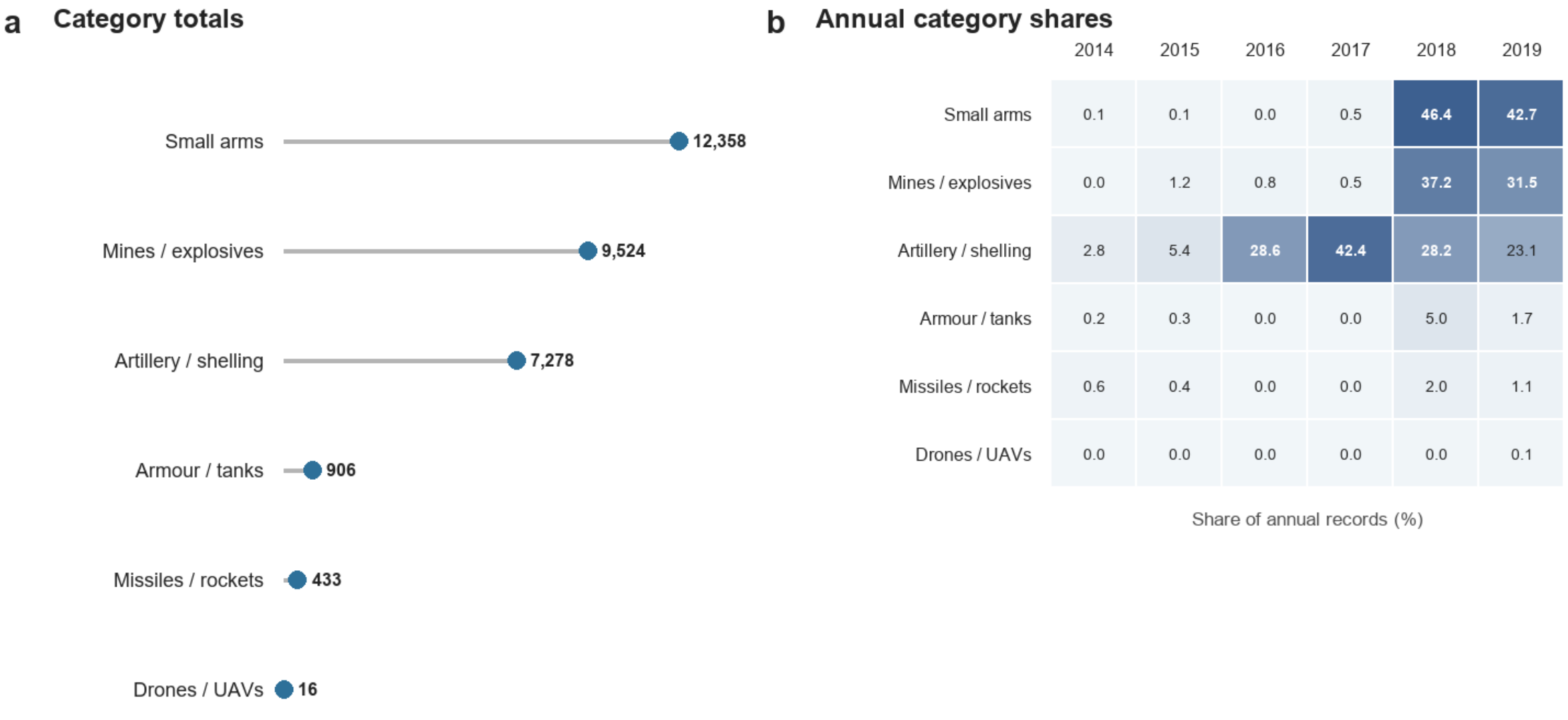

**Figure C1. Attack-category totals and annual composition of Donbas conflict records.**

# Appendix D. Project-level and annual semantic outcomes

Each ETIS title-abstract document was transformed with the reference-corpus term-frequency inverse-document-frequency (TF-IDF) vocabulary and normalized to unit Euclidean length (Salton & Buckley, 1988). A signed score was obtained by projecting the project vector onto the normalized difference between the positive and matched-comparison centroids, following the contrastive centroid logic developed by Rocchio (1971). Project scores were aggregated by calendar year as means, medians, top-decile shares, and positive-financing-weighted means. Capability maximum was calculated at the project level before annual aggregation, while entropy was calculated from the annual nonnegative shares of the eight capability channels.

Table D1 summarizes the mean, standard deviation, minimum, and maximum of the annual explicit-war, dual-use, and maximum-capability scores across 2000 to 2019.

**Table D1. Distribution of annual semantic outcomes.**

| Measure | Mean | SD | Minimum | Maximum |
|---|---|---|---|---|
| dual_use_mean | 0.0085 | 0.0016 | 0.0056 | 0.0108 |
| capability_max_mean | 0.0232 | 0.0041 | 0.0174 | 0.0312 |
| explicit_war_mean | 0.0032 | 0.0026 | -0.0034 | 0.0072 |

Figure D1 complements the distributional statistics in Table D1 by tracing the year-specific dual-use mean and marking the 2014 breakpoint.

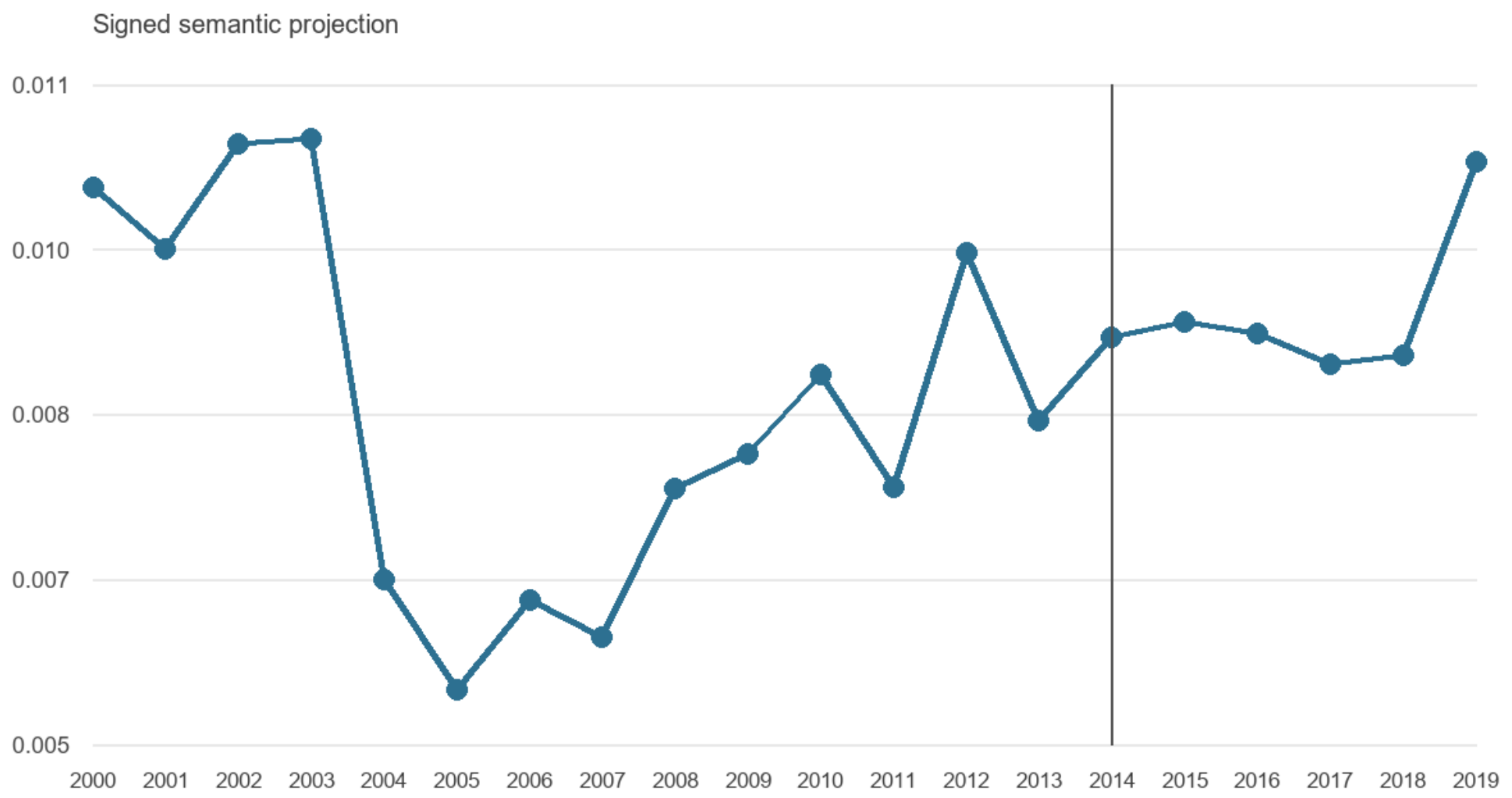

**Figure D1. Annual trajectory of the dual-use semantic mean.**

Figure D1 shows a high early-period level followed by a decline in 2004 and a gradual recovery thereafter. The series rises again after the 2014 breakpoint and reaches its post-breakpoint maximum in 2019. This trajectory is consistent with the positive post-2014 dual-use slope change reported in the main segmented estimates. Table D1 places the year-specific movement within the full observed range of the three principal annual semantic outcomes.

# Appendix E. Dual-use technology as a semantic construct

Dual-use technology describes knowledge, components, and systems whose functions span civilian and defense or security applications (Alic, 1994; Brandt, 1994). The concept is relational because use depends on technical affordances, organizational settings, and the problems to which a technology is applied. Contemporary work on artificial intelligence (AI) shows that diffusion between civilian and defense research and development (R&D) can occur through shared methods, personnel, infrastructures, and evaluation practices (Riebe et al., 2023). The present semantic measure therefore emphasizes application-relevant technical language rather than institutional labels alone.

Table E1 maps the eight capability domains to recurrent dual-use functions identified in defense-innovation and resilience literature.

**Table E1. Dual-use capability domains and application logic.**

| Domain | Illustrative functions | Basis |
|---|---|---|
| Engineering | Autonomy, control, infrastructure, logistics | (Mowery, 2012; North Atlantic Treaty Organization, 2020) |
| Materials | Protection, sensing, advanced manufacturing | (Alic, 1994; North Atlantic Treaty Organization, 2020) |
| Computer science | AI, cyber, communications, data | (North Atlantic Treaty Organization, 2020; Riebe et al., 2023) |
| Energy | Resilient supply, storage, propulsion | (European Commission, 2022; North Atlantic Treaty Organization, 2020) |
| Medicine | Trauma, emergency care, human performance | (North Atlantic Treaty Organization, 2020) |
| Life science | Biotechnology, detection, biosecurity | (Alic, 1994; North Atlantic Treaty Organization, 2020) |
| Earth and environment | Observation, geospatial sensing, hazards | (North Atlantic Treaty Organization, 2020) |
| Agriculture and food | Supply continuity, food-system resilience | (European Commission, 2022) |

# Appendix F. Complete exploratory event-semantic correlations

Spearman rank correlations were calculated between conflict-event series and contemporaneous or one-year-lagged annual semantic outcomes for 2014 to 2019. Exact two-sided probabilities were obtained by complete enumeration of the available annual ordering, following permutation-based exact inference (Ernst, 2004). Table F1 reports every tested event-outcome combination. The strongest lagged result links total conflict records to the following year's dual-use mean.

**Table F1. Complete event-semantic rank correlations.**

| outcome | exposure | lag | rho | exact_p | n |
|---|---|---|---|---|---|
| dual_use_mean | events | 0 | 0.3140 | 0.5640 | 6 |
| dual_use_mean | events | 1 | 1.0000 | 0.0250 | 5 |
| dual_use_mean | fatalities | 0 | -0.2000 | 0.7140 | 6 |
| dual_use_mean | fatalities | 1 | 0.7000 | 0.2400 | 5 |
| dual_use_mean | remote_violence | 0 | -0.2570 | 0.6590 | 6 |
| dual_use_mean | remote_violence | 1 | 0.7000 | 0.2400 | 5 |
| dual_use_mean | civilian_violence | 0 | 0.3380 | 0.5340 | 6 |
| dual_use_mean | civilian_violence | 1 | 0.7070 | 0.4050 | 5 |
| dual_use_top10_share | events | 0 | 0.2570 | 0.6590 | 6 |
| dual_use_top10_share | events | 1 | 0.7000 | 0.2400 | 5 |
| dual_use_top10_share | fatalities | 0 | -0.7710 | 0.1040 | 6 |
| dual_use_top10_share | fatalities | 1 | 0.3000 | 0.6860 | 5 |
| dual_use_top10_share | remote_violence | 0 | 0.1430 | 0.8030 | 6 |
| dual_use_top10_share | remote_violence | 1 | 0.7000 | 0.2400 | 5 |
| dual_use_top10_share | civilian_violence | 0 | 0.6760 | 0.2010 | 6 |
| dual_use_top10_share | civilian_violence | 1 | 0.7070 | 0.4050 | 5 |
| capability_max_mean | events | 0 | 0.4860 | 0.3560 | 6 |
| capability_max_mean | events | 1 | 0.6000 | 0.3550 | 5 |
| capability_max_mean | fatalities | 0 | -0.6000 | 0.2430 | 6 |
| capability_max_mean | fatalities | 1 | 0.1000 | 0.9500 | 5 |
| capability_max_mean | remote_violence | 0 | 0.4290 | 0.4200 | 6 |
| capability_max_mean | remote_violence | 1 | 0.9000 | 0.0910 | 5 |
| capability_max_mean | civilian_violence | 0 | 0.8450 | 0.0680 | 6 |
| capability_max_mean | civilian_violence | 1 | 0.7070 | 0.4050 | 5 |
| capability_max_top10_share | events | 0 | 0.4860 | 0.3560 | 6 |
| capability_max_top10_share | events | 1 | 0.6000 | 0.3550 | 5 |
| capability_max_top10_share | fatalities | 0 | -0.6000 | 0.2430 | 6 |
| capability_max_top10_share | fatalities | 1 | 0.1000 | 0.9500 | 5 |
| capability_max_top10_share | remote_violence | 0 | 0.4290 | 0.4200 | 6 |
| capability_max_top10_share | remote_violence | 1 | 0.9000 | 0.0910 | 5 |
| capability_max_top10_share | civilian_violence | 0 | 0.8450 | 0.0680 | 6 |
| capability_max_top10_share | civilian_violence | 1 | 0.7070 | 0.4050 | 5 |

# References for the Appendices

Alic, J. A. (1994). The dual use of technology: Concepts and policies. *Technology in Society*, 16(2), 155–172. https://doi.org/10.1016/0160-791X(94)90027-2

Brandt, L. (1994). Defense conversion and dual-use technology. *Policy Studies Journal*, 22(2), 359–370. https://doi.org/10.1111/j.1541-0072.1994.tb01474.x

Ernst, M. D. (2004). Permutation methods: A basis for exact inference. *Statistical Science*, 19(4), 676–685. https://doi.org/10.1214/088342304000000396

European Commission. (2022). *EU research and innovation and the invasion of Ukraine: Main channels of impact*. https://research-and-innovation.ec.europa.eu/knowledge-publications-tools-and-data/publications/all-publications/eu-research-and-innovation-and-invasion-ukraine-main-channels-impact_en

Mowery, D. C. (2012). Defense-related R&D as a model for grand challenges technology policies. *Research Policy*, 41(10), 1703–1715. https://doi.org/10.1016/j.respol.2012.03.027

North Atlantic Treaty Organization. (2020). *Science & Technology Trends 2020–2040: Exploring the S&T Edge*. https://www.nato.int/nato_static_fl2014/assets/pdf/2020/4/pdf/190422-ST_Tech_Trends_Report_2020-2040.pdf

Priem, J., Piwowar, H., & Orr, R. (2022). OpenAlex: A fully-open index of scholarly works, authors, venues, institutions, and concepts. *Quantitative Science Studies*, 3(3), 687–708. https://doi.org/10.1162/qss_a_00197

Raleigh, C., Linke, A., Hegre, H., & Karlsen, J. (2010). Introducing ACLED: An armed conflict location and event dataset. *Journal of Peace Research*, 47(5), 651–660. https://doi.org/10.1177/0022343310378914

Riebe, T., Schmid, S., & Reuter, C. (2023). Dual-Use and Trustworthy? A Mixed Methods Analysis of AI Diffusion Between Civilian and Defense R&D. *Science and Engineering Ethics*, 29, 18. https://doi.org/10.1007/s11948-023-00431-1

Rocchio, J. J. (1971). Relevance feedback in information retrieval. In G. Salton (Ed.), *The SMART Retrieval System: Experiments in Automatic Document Processing* (pp. 313–323). Prentice-Hall.

Salton, G., & Buckley, C. (1988). Term-weighting approaches in automatic text retrieval. *Information Processing & Management*, 24(5), 513–523. https://doi.org/10.1016/0306-4573(88)90021-0

Sundberg, R., & Melander, E. (2013). Introducing the UCDP Georeferenced Event Dataset. *Journal of Peace Research*, 50(4), 523–532. https://doi.org/10.1177/0022343313484347